\documentclass[12pt]{article}
\usepackage{amsmath,amssymb}
\usepackage{epsfig}

\newcommand{\nn}{\nonumber}

\newcommand{\hs}{\hat{s}}

\newcommand{\qq}{q\bar{q}}
\newcommand{\ttb}{t\bar{t}}

\newcommand{\bea}{\begin{eqnarray}}
\newcommand{\eea}{\end{eqnarray}}
\newcommand{\simgt}
{\hbox{ \raise3pt\hbox to 0pt{$>$}\raise-3pt\hbox{$\sim$} }}
\newcommand{\simlt}
{\hbox{ \raise3pt\hbox to 0pt{$<$}\raise-3pt\hbox{$\sim$} }}

\begin{titlepage}

\title{
Bound-state Effects on Top Quark Production\\ at Hadron Colliders}

\author{
{}\\
Kaoru Hagiwara\footnote{E-mail address:
 \texttt{kaoru.hagiwara@kek.jp}}\\
 {\normalsize \sl KEK Theory Division and Sokendai, Tsukuba 305-0801,
 JAPAN}\\ \\
Yukinari Sumino\footnote{E-mail address:
 \texttt{sumino@tuhep.phys.tohoku.ac.jp}}\\
{\normalsize \sl Dept.\ of Physics, Tohoku University, Sendai 980-8578,
 JAPAN}\\ \\
Hiroshi Yokoya\footnote{E-mail address:
           \texttt{hiroshi.yokoya@cern.ch}}\\
{\normalsize \sl Theory Unit, Physics Department, CERN, CH-1211 Geneva,
 SWITZERLAND}
}
\date{}
\end{titlepage}

\begin{document}

\thispagestyle{empty}

\maketitle
\abstract{
We study bound-state effects on the $\ttb$ production cross section in
the threshold region at hadron colliders.
The bound-state effects are important
particularly at the LHC where the gluon fusion is the dominant
subprocess.
Due to the formation of $\ttb$ resonances in the $J=0$ color--singlet
channel of $gg\to\ttb$ and the large width of the top quark, the $\ttb$
invariant-mass distribution peaks at a few GeV below the $\ttb$ threshold,
and it is significantly enhanced over the naive NLO prediction until
several GeV above the threshold.
We present predictions of the $\ttb$ invariant-mass distribution which
incorporate both the bound-state effects and initial-state radiations up
to NLO.
The bound-state effects would lead to a substantial deformation of
top-quark kinematical distributions in the threshold region.
}
\vspace{-18.5cm}
\begin{flushright}
{KEK-TH-1231}\\
{TU-811}\\
{CERN-PH-TH-2008-045}\\
{April 2008}
\end{flushright}

\newpage

% Introduction

In the forthcoming CERN Large Hadron Collider (LHC) experiment, top
quarks will be produced copiously.
A huge top-quark pair production event sample is considered as a good
template for performing various physics studies as well as detector
calibrations.
For this purpose, it is important that we understand
physics of the top-quark production and decay accurately.
Through dedicated studies of the top-quark pair production events
at the Fermilab Tevatron, it has been recognized that top-quark events
can be reconstructed with good accuracy.
For instance, using a top-quark sample in lepton+4-jet mode,
the lepton helicity-angle distribution in top-quark decays
has been measured,
which requires a full kinematical reconstruction of each event.
Up to now, a good agreement with the Standard-Model
prediction has been observed for this distribution
\cite{Abulencia:2006ei}.

At the LHC, top quarks are produced dominantly via gluon-gluon fusion.
The gluon distribution function is a rapidly decreasing function of its
momentum fraction $x$.
Thus, a substantial amount of top quarks are expected to be
pair-produced close to their production
threshold in gluon-gluon fusion at the LHC.
Therefore, contributions of $\ttb$ resonances may be important
in the threshold region.
The dominant contribution stems from the $J^{PC}=0^{-+}$ ($L=S=0$)
color--singlet $\ttb$ resonance states.
An effective operator for producing one of these resonances may be
written as
\bea
{\cal L}_{gg\to\phi(\ttb)}\propto
\epsilon^{\mu\nu\rho\sigma}G^a_{\mu\nu}G^a_{\rho\sigma}\,
\phi(\ttb) .
\eea
In contrast, at the Tevatron, the quark-antiquark fusion
is the main process of top production.
Hence, the $\ttb$ pairs are predominantly in the color--octet $J=1$
state, and we anticipate that resonance effects are not significant.

The method for incorporating $\ttb$ bound-state effects has
been developed mainly
in the studies of $\ttb$ productions in
$e^+e^-$ collisions~\cite{Fadin:1987wz,Hoang:2000yr}.
Formally, in the limit where we neglect the top-quark width,
$\Gamma_t\to0$, bound-state effects can be incorporated by
resummation of the threshold singularities
$(\alpha_s/\beta)^n$, where 
$\beta$ 
is the velocity of
the top quark in the $\ttb$ c.m.\ frame.
It is known that, due to the large top-quark width, resonance peaks
in the top-quark pair production cross section are smeared out, and a
broad enhancement of the cross section over the entire threshold region
results from the bound-state effects.

On the other hand,
a distinct feature of $\ttb$ productions in hadron collisions 
is that effects of
initial-state radiation (ISR) of gluons are significant, especially
close to the top pair production threshold
\cite{Nason:1987xz,Beenakker:1988bq,Bonciani:1998vc}.
They can be incorporated by resummation of soft and collinear 
logarithms, which are given in the form $(\alpha_s \ln^2 \beta)^n$.
So far, up to our knowledge, only two papers
\cite{Fadin:1990wx,Catani:1996dj}
studied bound-state effects on $\ttb$ productions at hadron
colliders that are relevant for a realistic top-quark mass.\footnote{
Earlier works \cite{Pancheri:1992km,Kuhn:1992qw} considered
production of sharp toponium resonances.
Their phenomenology is rather different from that of 
today's realistic $\ttb$ resonances, which have much larger decay widths.
}
In Ref.~\cite{Fadin:1990wx}, the $\ttb$ invariant-mass distributions
are computed in the threshold region
incorporating the leading-order (LO) bound-state
effects (Coulomb effects) and ignoring other QCD corrections.
In Ref.~\cite{Catani:1996dj},
in the study of the total production cross sections of top quarks,
the Coulomb effects above the $\ttb$ threshold are examined in LO,
in comparison to the
next-to-leading order (NLO) ISR effects.

In this paper, we study the $\ttb$ invariant-mass
distributions in the threshold region,
incorporating both bound-state effects and ISR effects
at NLO.
The differences of the present work from the
previous ones are as follows.
As compared to Ref.~\cite{Fadin:1990wx}, 
we incorporate 
the ISR effects and the NLO corrections to the
bound-state effects.
As compared to Ref.~\cite{Catani:1996dj},
we examine the $\ttb$ invariant-mass distributions;
we include contributions from the 
resonances below the $\ttb$ threshold;
we include the NLO corrections to the
bound-state effects.\\

In the computation of the parton-level cross sections,
the bound-state effects and the ISR
effects factorize in the LO of
the threshold singularities and in the
leading-logarithmic (LL) approximation
of the soft+collinear 
singularities~\cite{Bodwin:1994jh,Bonciani:1998vc}:
\bea
\hat\sigma_{\rm ISR}(\hs ; i\to f) =
K_i^{(c)}\, \int_0^1 dz\, \, \hat\sigma(s'=z\hat{s};i\to f)
\, F^{(c)}_{i}(z) .
\label{eq:convol}
\eea
Here, $\hat\sigma(s';i\to f)$ is the cross section without ISR
($i=q\bar{q}$ or $gg$, $f=\ttb$ in various color and $J$
states); $\hat\sigma_{\rm ISR} (\hs ; i\to f)$ is the cross section
including the ISR effects.
The ISR function $F^{(c)}_i$ depends on the initial-state
partons ($i$) and the color of the final-state $\ttb$ system ($c=1$ and
$8$ for color--singlet and octet $\ttb$ states, respectively).
$K_i^{(c)}$ denotes the hard-vertex correction factor,
which is normalized as $K_i^{(c)}=1$ in the leading order.

In this paper, for simplicity of our analysis, we include the ISR
effects up to NLO.
It is known that the differences between the NLO and next-to-leading log
(NLL) corrections to the $\ttb$ production cross sections are small,
if we ignore bound-state effects \cite{Catani:1996dj,Bonciani:1998vc}.
Hence, the NLO approximation may be justified
for our first analysis of the threshold cross sections.
The ISR functions for the production of various color states
up to NLO are given by~\cite{Kuhn:1992qw,Petrelli:1997ge}
\bea
&&
F^{(c)}_{i}(z)=\delta{(1-z)} + \frac{\alpha_s(\mu_F)}{\pi}
\left[f_{i}^{(c)}\left(z,\frac{\mu_F}{2m_t}\right)
+k_i^{(c)}\left(\frac{\mu_F}{2m_t}\right)\,\delta(1-z)
\right] .
\label{ISRfn}
\eea
The ${\cal O}(\alpha_s)$ terms read\footnote{
The leading (double-log) terms $\propto [\ln(1-z)/(1-z)]_+$ of
$f_{gg}^{(c)}$ and $f_{\qq}^{(8)}$ are common to those of the
pseudo-scalar Higgs boson production via gluon
fusion process~\cite{Spira:1993bb} and 
the Drell-Yan process~\cite{Altarelli:1979ub}, respectively.
}
\begin{subequations}
\bea
&& f_{gg}^{(1)}\left(z,\frac{\mu_F}{2m_t}\right)
=4C_A\left[\left(\frac{\ln{(1-z)}}{1-z}\right)_{+}
-\left(\frac{1}{1-z}\right)_{+}\ln{\left(\frac{\mu_F}{2m_t}\right)}\right],
\\ &&
 f_{gg}^{(8)}\left(z,\frac{\mu_F}{2m_t}\right)
=4C_A\left[\left(\frac{\ln{(1-z)}}{1-z}\right)_{+}
-\left(\frac{1}{1-z}\right)_{+}\ln{\left(\frac{\mu_F}{2m_t}\right)}\right]
\nonumber\\&&
~~~~~~~~~~~~~~~~~~~~~
 -C_A\left(\frac{1}{1-z}\right)_{+},
\\&&
 f_{\qq}^{(8)}\left(z,\frac{\mu_F}{2m_t}\right)
=4C_F\left[\left(\frac{\ln{(1-z)}}{1-z}\right)_{+}
-\left(\frac{1}{1-z}\right)_{+}\ln{\left(\frac{\mu_F}{2m_t}\right)}\right]
\nonumber\\&&
~~~~~~~~~~~~~~~~~~~~~
 -C_A\left(\frac{1}{1-z}\right)_{+},
\label{eq:sgr}
\eea
\end{subequations}
\begin{subequations}
\bea
&&k_{gg}^{(c)} \left(\frac{\mu_F}{2m_t}\right) =
-\beta_0\ln{\left(\frac{\mu_F}{2m_t}\right)},\\
&&k_{\qq}^{(c)}\left(\frac{\mu_F}{2m_t}\right) =
-3C_F\ln{\left(\frac{\mu_F}{2m_t}\right)},
\eea\label{subeq:k}
\end{subequations}
\noindent
\!\!with $\beta_0 = \frac{11}{3}C_A-\frac{2}{3}n_q$.
Here, $m_t$ and $\mu_F$ denote
the top-quark pole mass and the factorization scale, respectively
(we take $\mu_F=m_t$);
$\alpha_s$ is the strong coupling constant in the
$\overline{\rm MS}$ scheme;
$C_F=4/3$ and $C_A=3$ are color factors,
and we take the number of light quark flavors to be $n_q=5$.
The plus-distribution is defined in the standard manner.

Let us turn to the threshold cross sections without ISR,
$\hat\sigma(s';i\to f)$.
As is well known, the $S$-wave part of the cross sections are
most important in the threshold region.
Contributions of $L>0$ are suppressed at least by 
$\beta^2\sim\alpha_s^2$
with respect to the leading $S$-wave contribution.
For individual cross sections, leading $S$-wave contributions reside in
the following channels:
\begin{subequations}
\bea
&i=gg,
&~f=\ttb(L=0,~S=0,~J=0,~\mbox{color singlet}) ;
\\
&i=gg,
&~f=\ttb(L=0,~S=0,~J=0,~\mbox{color octet} );
\\
&i=q\bar{q},
&~f=\ttb(L=0,~S=1,~J=1,~\mbox{color octet} ).
\eea
\end{subequations}
There is no color--singlet channel for $i=q\bar{q}$ at LO.
The process $gg \to \ttb (L=0,S=J=1)$ in the color--singlet channel
is forbidden due to the angular momentum conservation and Bose
statistics (Yang's theorem), and the same applies to the symmetric
($d^{abc}$) part of the color-octet channel.
It is also
forbidden in the anti-symmetric ($f^{abc}$) part of the
color--octet channel at LO;
this is because it is
a na\"ive-$T$ odd transition and is hence forbidden at LO due to the
time-reversal invariance of QCD.
We ignore $qg$ subprocess contributions, which are suppressed
by $\alpha_s\beta^2$ as compared to the LO contribution.

Following the standard framework developed for threshold cross
sections \cite{Fadin:1987wz,Hoang:2000yr},
the $S$-wave cross section for $i \to f$ including bound-state effects
is given by
\bea
\hat\sigma(s';i\to f) =
\left[\hat\sigma(s';i\to f) \right]_{\rm tree}\times
\frac{{\rm Im}[G^{(c)}(\vec{0};E)]}
{{\rm Im}[G_0(\vec{0};E)]} .
\label{inclBSeff}
\eea
The non-relativistic Green functions are defined by
\bea
\left[ (E+i\Gamma_t)-
\left\{ - \frac{\nabla^2}{m_t} + V_{\rm QCD}^{(c)}(r)\right\}\right]
G^{(c)}(\vec{x};E) = \delta^3(\vec{x}) ,
\label{defGreenfn}
\eea
where $E=\sqrt{s'}-2m_t$ is the c.m.\ energy of the $\ttb$ system
measured from the threshold; $r=|\vec{x}|$;
$\Gamma_t$ is the decay width of the top quark;
$V_{\rm QCD}^{(c)}(r)$
is the QCD potential between the
color--singlet ($c=1$) or color--octet ($c=8$)
static quark-antiquark pair.
On the other hand,
$G_0(\vec{x};E)$ is the non-relativistic Green function of
a free $\ttb$ pair, which is defined via Eq.~(\ref{defGreenfn})
after setting $V_{\rm QCD}^{(c)}(r)$ and $\Gamma_t$ to zero.

We use the NLO QCD potential, which reads \cite{Kniehl:2004rk}:
\bea
V_{\rm QCD}^{(c)}(r;\mu_{\rm B})=
C^{(c)} \frac{\alpha_s(\mu_{\rm B})}{r}
\left[
1 + \frac{\alpha_s(\mu_{\rm B})}{4\pi}
\left\{ 2\beta_0\left[\ln(\mu_{\rm B} r) + \gamma_E\right]+ a_1^{(c)}
\right\}
\right]
\eea
with
\bea
&&
C^{(1)}=-C_F,
~~~~
C^{(8)}=\frac{C_A}{2}-C_F ,
\\&&
a_1^{(1)}=a_1^{(8)}= \frac{31}{9}C_A - \frac{10}{9}n_q,
\eea
for the $\overline{\rm MS}$ coupling.
Here, $\gamma_E=0.5772...$ denotes the Euler constant.
The QCD potential is renormalization-group invariant,
and we evaluate the above expression at the Bohr scale of
$\mu_{\rm B}=20$~GeV, and with $n_q=5$.

Eq.~(\ref{inclBSeff}) incorporates the QCD
bound-state effects between $\ttb$ up to NLO.
It also incorporates the top-quark decay-width effects 
on the cross section $\hat\sigma(s';i\to f) $ up to NLO,
provided that the ${\cal O}(\alpha_s)$ correction
to the top-quark decay width is
included in $\Gamma_t$
\cite{Fadin:1987wz,Melnikov:1993np}.

We correct the
$gg\to\ttb$ tree-level cross sections in the $J=0$ color--singlet
and octet channels using
Eq.~(\ref{inclBSeff}), since they include all the leading $S$-wave
contributions;
we do not include bound-state effects in the $J>0$ channels.
In the $q\bar{q}\to\ttb$ tree-level cross section, there is only
color--octet $J=1$ channel which contains the $S$-wave contribution, and
hence we correct it by using the octet Green function via
Eq.~(\ref{inclBSeff}).
\setcounter{footnote}{0}

The $S$-wave Born cross sections behave as $\hat{\sigma}\propto
\alpha_s^2\beta$ near the threshold ($\beta\ll 1$).
Expanding $\hat{\sigma}_{\rm ISR}$ [Eq.~(\ref{eq:convol})] in $\alpha_s$, we
correctly reproduce the dominant
${\cal O}(\alpha_s^3)$ corrections of
the NLO cross sections \cite{Nason:1987xz}, namely,
$\alpha_s^3\beta\ln^2{\beta}$,
$\alpha_s^3\beta\ln{\beta}$ and $\alpha_s^3\beta^0$ terms.
Furthermore,  by introducing the following
hard-vertex correction factors in Eq.~(\ref{eq:convol}),
we match $\hat{\sigma}_{\rm ISR}$
to the NLO cross sections 
up to the $\alpha_s^3\beta$ term:
\bea
K_{i}^{(c)}= 1 + \frac{\alpha_s(\mu_R)}{\pi}\,
h_i^{(c)}\left(\frac{\mu_R}{m_t}\right)
\label{factork}
\eea
with
\begin{subequations}
\begin{align}
h_{gg}^{(1)}\left(\frac{\mu_R}{m_t}\right) & =
 C_A\left(1+\frac{\pi^2}{12}\right) +C_F\left(-5+\frac{\pi^2}{4}\right)
+ \beta_0\ln{\left(\frac{\mu_R}{2m_t}\right)},\\
h_{gg}^{(8)}\left(\frac{\mu_R}{m_t}\right) & =
 C_A\left(3-\frac{\pi^2}{24}\right) +C_F\left(-5+\frac{\pi^2}{4}\right)
+ \beta_0\ln{\left(\frac{\mu_R}{2m_t}\right)},\\
h_{\qq}^{(8)}\left(\frac{\mu_R}{m_t}\right) & =
C_A\left(\frac{59}{9}-\frac{\pi^2}{4}+\frac{2\ln{2}}{3}\right)
+C_F\left(-8+\frac{\pi^2}{3}\right)\nn\\
&\quad -\frac{5}{9}n_q -\frac{8}{9}
+\beta_0\ln{\left(\frac{\mu_R}{2m_t}\right)}.
\end{align}\label{subeq:h}
\end{subequations}
\noindent
\!\!Here, $\mu_R$ denotes the renormalization scale in the
$\overline{\rm MS}$ scheme.
(We take $\mu_R=m_t$.)
The logarithmic part of $h_i^{(c)}$  in Eqs.~(\ref{subeq:h}) 
are independent of the color of
the $\ttb$ system.
They cancel the renormalization-scale dependence of the Born
cross sections which are proportional to $\alpha_s^2(\mu_R)$.
The non-logarithmic part of $h_{i}^{(c)}$ are
extracted from the NLO cross sections for the heavy quarkonium
productions;
$h^{(1)}_{gg}(1)$ from the results of
Refs.~\cite{Kuhn:1992qw,Petrelli:1997ge};
$h^{(8)}_{gg}(1)$ and $h^{(8)}_{\qq}(1)$
from the results of Ref.~\cite{Petrelli:1997ge}.
Numerically they read  $h^{(1)}_{gg}(1)\approx -3.22$,
$h^{(8)}_{gg}(1)\approx -0.92$, and $h^{(8)}_{\qq}(1)\approx -1.61$.
We note that
the term $-{8}/{9}$ in $h_{\qq}^{(8)}$
is missing in Ref.~\cite{Petrelli:1997ge}.\footnote{
The non-decoupling effect of the
top-quark loop in the gluon vacuum polarization in the $s$-channel 
diagram contributes 
$    2[ \Pi_t(s=4 m_t^2) - \Pi_t(s=0) ] = - 8/9$,
where $\Pi_t$ is the top quark contribution to the gluon two-point
function.
The factor of 2 appears since it is a part of the interference 
terms.
}
This is due to the fact that 
non-decoupling effects from 
heavy-quark loops 
are incorrectly omitted
throughout the computations in Ref.~\cite{Petrelli:1997ge}.\footnote{%
We thank the authors of
Ref.~\cite{Petrelli:1997ge} for confirming this error in their results.
We also thank M.~Czakon and A.~Mitov for providing the analytic
expression of the $\alpha_s^3\beta$ term of the
$\qq\to\ttb$ cross section, prior to its publication~\cite{Czakon},
which was crucial for us to identify the source of the discrepancy
between Refs.~\cite{Nason:1987xz} and \cite{Petrelli:1997ge}.
}

We can extract
$h^{(8)}_{\qq}(1)$ numerically also from the NLO $\qq\to\ttb$
cross section \cite{Nason:1987xz}, which
reads $h^{(8)}_{\qq}(1)\approx -1.8$.
The agreement is fairly good.
Similarly, we can extract a color-weighted sum of $h^{(c)}_{gg}(1)$
from the NLO $gg\to\ttb$ cross section \cite{Nason:1987xz} as
$\frac{2}{7}h_{gg}^{(1)}(1)+\frac{5}{7}h_{gg}^{(8)}(1)\approx 0.8$, which
is quite different from the corresponding value $-1.58$ that follows from
Eqs.~(\ref{subeq:h}).
Nevertheless, we 
confirmed that these two rather different values are marginally
consistent with each other if we take into account
the claimed numerical accuracy of Ref.~\cite{Nason:1987xz}.\footnote{%
A precise statement is as follows.
The present discrepancy corresponds to a 7\% difference in terms of
$a_0$, which is the only parameter contributing to the 
$\alpha_s^3\beta$ term in the fitting function in
Ref.~\cite{Nason:1987xz}.
We have checked that it is possible to shift the value of $a_0$ by 7\%
without altering the fitting function more than 1\%,
which is the accuracy claimed for the fit, if we adjust
the remaining parameters appropriately.}
We will discuss this issue further at the end of the paper.

In general, non-logarithmic parts of $k_i^{(c)}$ and $h_i^{(c)}$
both contribute to the $\alpha_s^3\beta$ term of the cross section.
How to separate non-logarithmic parts of $k_i^{(c)}$ and $h_i^{(c)}$ is
scheme dependent.
In this paper, we choose a scheme such that non-logarithmic part of
$k_i^{(c)}$ is zero; see Eqs.~(\ref{subeq:k}).

Finally, convoluting with the parton distribution functions (PDFs), we
obtain the hadronic cross section
\begin{align}
\sigma(s;i\to f)&=\int_{0}^{1}d\tau\,
 \frac{d{\cal L}_{i}}{d\tau}(\tau)\,
\hat\sigma_{\rm ISR}(\hat{s}=\tau s;i\to f)\label{eq:tot} \\
&=K^{(c)}_{i}\int_{0}^{1}d\tau\,
 \frac{d{\cal L}_{i}}{d\tau}(\tau)\,
\int_{0}^1dz\, \hat\sigma(z\tau s;i\to f)\,F_{i}^{(c)}(z).
\end{align}
The partonic luminosity is defined by
\begin{align}
 \frac{d{\cal L}_{i}}{d\tau} (\tau;\mu_F) =
\sum_{\{a,b\}}\int dx_1\int dx_2\,f_{a}(x_1,\mu_F)\,f_{b}(x_2,\mu_F)
\,\delta{(\tau-x_1 x_2)},\label{eq:lum}
\end{align}
 where the summation is over $\{a,b\}=\{g,g\}$ for $i=gg$, and
 $\{a,b\}=\{q,\bar{q}\},\{\bar{q},q\}$ with $q=\{u,d,s,c,b\}$ for
 $i=\qq$.
In numerical calculations, we use the CTEQ6M (NLO, standard
 $\overline{\rm MS}$ scheme) PDF parametrization~\cite{Pumplin:2002vw}
 at the factorization scale of $\mu_F=m_t$.
We present our results in the form of the $\ttb$ invariant-mass
distribution, which is, in principle, a measurable quantity
from the final-states at hadron colliders:
\bea
&&
\frac{d\sigma}{dm_{\ttb}}(s,m^2_{\ttb};i\to f)
= \frac{2m_{\ttb}}{s} \,
\hat\sigma(m_{\ttb}^2\,;i\to
 f) \times K^{(c)}_{i}
\int^1_{\tau_0}\frac{dz}{z}\, F_{i}^{(c)}(z)\,
\frac{d{\cal L}_i}{d\tau}\!\left(\frac{\tau_0}{z}\right) .
\nonumber\\
\label{eq:mtt}
\eea
Here, $m_{\ttb}$ denotes the invariant-mass of $\ttb$,
and $\tau_0=m^2_{\ttb}/s$.
The ISR function $F^{(c)}_{i}(z)$ is convoluted with the
 partonic luminosity but not with $\hat{\sigma}$, which 
is evaluated at fixed $s'=m_{\ttb}^2$.

Our formulas for $\hat{\sigma}(s';i\to f)$ rely on the
non-relativistic QCD framework \cite{Hoang:2000yr} 
and are valid only in the threshold region.
More specifically,
our formulas are subject to ${\cal O}(\alpha_s^3 \beta^2)$ corrections
that grow with $E$
(part of NNLO corrections).
As a result, the $\ttb$ invariant-mass
distributions described above are valid only
in the threshold region  $\beta\ll 1$;
they approach the NLO results of Ref.~\cite{Frederix:2007gi}
in the region $\alpha_s\ll\beta \ll 1$
but deviate from the NLO results 
at higher invariant-mass $\beta \sim {\cal O}(1)$.
Hence, our predictions
should be smoothly interpolated to the corresponding
NLO results in the region 
$\alpha_s\ll\beta \ll 1$.
Numerically smooth interpolations may be performed
at $m_{\ttb}-2m_t \sim$~5--20~GeV.\footnote{
In this regard, 
we note that the ratio of the Green functions
in Eq.~(\ref{inclBSeff}) has the following high energy
behavior:
\bea
\frac{{\rm Im}\,G^{(c)}}{{\rm Im}\,G_0}
\to
1 - \frac{\pi}{2}\,C^{(c)}\,\alpha_s(\mu_B)
+ {\cal O}(\alpha_s^2)
~~~\mbox{for}
~~~
E \gg 2m_t~~
(\beta \to 1).
\nonumber
\eea
}
\\

Below we examine the $\ttb$ invariant-mass distributions
numerically.
We compare the invariant-mass distributions
which include QCD corrections in four different
ways:\vspace{3mm}\\
\begin{tabular}{lp{105mm}}
Born:& The distribution at the Born level.
On the right-hand side of Eq.~(\ref{eq:mtt}), 
$\hat{\sigma}$ is
replaced by the Born cross section;
the ISR function $F^{(c)}_{i}(z)$ 
is set to $\delta{(1-z)}$;
the hard-vertex factor $K_i^{(c)}$ is set to 1.
\\
NLO:&
The distribution with ISR effects but without bound-state
     effects.
The right-hand side of
Eq.~(\ref{eq:mtt}) 
(except ${d{\cal L}_i}/{d\tau}$)
is replaced by its expansion in $\alpha_s$
up to ${\cal O}(\alpha_s^3)$.
Note that it includes the $\alpha_s/\beta$ part of the
Coulomb-gluon exchange effects in ${\rm Im}G^{(c)}/{\rm Im}G_0$.
\\
Gr--Fnc:&
The distribution with bound-state effects but without ISR effects.
On the right-hand side of Eq.~(\ref{eq:mtt}), 
$F^{(c)}_{i}(z)$ is set to $\delta{(1-z)}$ and
$K_i^{(c)}$ to 1.
\\
Gr--Fnc$\times$ISR:&
Our full prediction, Eq.~(\ref{eq:mtt}).
\end{tabular}\vspace{3mm}

\begin{figure}[t]
\begin{center}
\epsfig{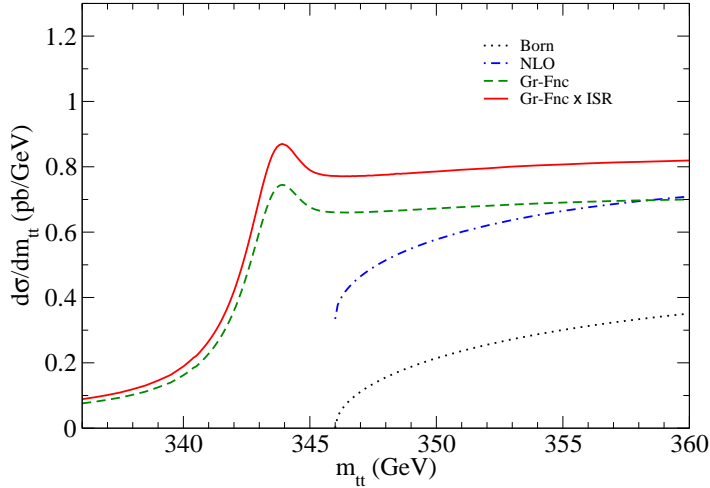}
\caption{
$\ttb$ invariant-mass distributions for $gg\to\ttb$ in
 the threshold region, in the $J=0$, color--singlet channel at the LHC.
The cross sections calculated in the Born approximation (black dotted),
 with bound-state effects but without ISR effects (green dashed), with
 ISR effects but without bound-state effects
 (blue dot-dashed), and with both bound-state and ISR effects (red solid)
 are plotted.
We take $m_t=173$~GeV and $\Gamma_t=1.5$~GeV.
}\label{fig:sngj0}
\end{center}
\end{figure}

In Fig.~\ref{fig:sngj0}, we plot the $\ttb$ invariant-mass distributions
for $gg \to \ttb$ in the $J=0$, color--singlet
channel at the LHC.
The above four cross sections are
shown (dotted, dot-dashed, dashed, and solid lines, respectively),
for $m_t=173$~GeV and $\Gamma_{t}=1.5$~GeV.
The effects of $\Gamma_t$ are included only through
the Green function in Eq.~(\ref{defGreenfn}).
The cross sections with binding effects exhibit a resonance
peak below the threshold, $m_{\ttb}=2m_t=346$~GeV.
Due to the large width of the top quark, multiple resonance peaks are
smeared out and only the broad $1S$ peak remains in $d\sigma/dm_{\ttb}$;
the feature well-known in the total cross section for $e^+e^-\to\ttb$.

The peak position is at $E= m_{\ttb}-2m_t=-2.1$~GeV, which
is very close to the peak position of the singlet Green function
${\rm Im}\,G^{(1)}$.
In fact, the rapid fall off of the partonic luminosity
$d{\cal L}_i/d\tau$ reduces the peak position by only about a few tens
MeV.
On the other hand, the ISR effects scarcely change the peak position.
This means that the peak positions of the $\ttb$ invariant-mass
distributions in the color--singlet channels
are, to a very good approximation, the same in $e^+e^-$ and
hadron collider experiments.

The ISR effects enhance the invariant-mass distribution obtained with
only the bound-state effects (Gr--Fnc)\
almost independently of $m_{\ttb}$, where the enhancement factor is
about $1.2$.
We also observe that the enhancement by ISR effects
is much more pronounced without the bound-state effects than after
inclusion of bound-state effects.
Namely, the ratio of the approximations NLO and Born is much larger than
that of Gr--Fnc$\times$ISR and Gr--Fnc.
This is because a substantial part of
the former enhancement comes
from the $\alpha_s/\beta$ term included in NLO,
which is genuinely a part of bound-state effects, whereas
this term is included in the Gr--Fnc approximation.

\begin{figure}[t]
\begin{center}
\epsfig{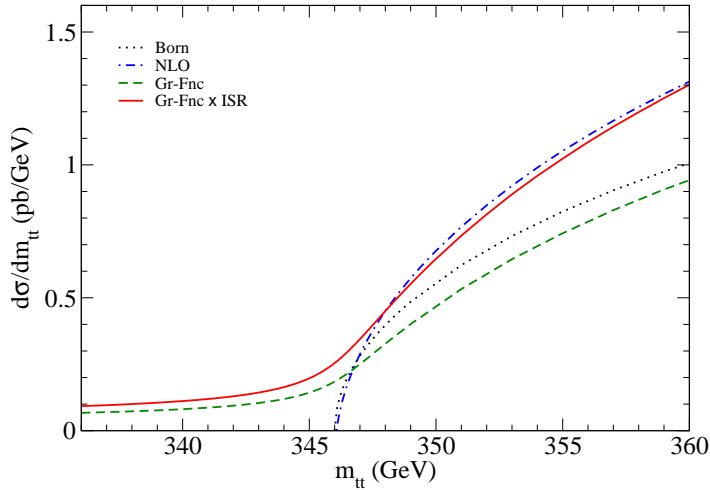}
\caption{
Same as Fig.~\ref{fig:sngj0}, but for $gg\to\ttb$ in the color--octet
 channel and all $J$ states are summed.}\label{fig:oct}
\end{center}
\end{figure}
Fig.~\ref{fig:oct} shows the invariant-mass distribution in the $gg$
 channel as in Fig.~\ref{fig:sngj0}, but for the color--octet states and
 when all $J$ contributions are summed.
We have included bound-state effects only in the $J=0$ channel.
By comparing Gr--Fnc$\times$ISR (solid) and NLO (dot-dashed),
one finds that the bound-state effects are not significant.
More precisely, among the bound-state effects $(\alpha_s/\beta)^n$,
contributions from $n\ge 2$ are rather small for the color--octet state.
A small tail of the cross section below the threshold
originates from the large width of the top quark.
The ratio of Gr--Fnc$\times$ISR and Gr--Fnc is
almost independent of $m_{\ttb}$ and about $1.4$.

The striking difference of the bound-state effects in the
color--singlet and color--octet channels can be
understood as follows.
The force between $t$ and $\bar{t}$ is attractive in the color--singlet
channel, leading to formation of resonance states below the threshold
$m_{\ttb}<2m_{t}$.
On the other hand, the interquark force is repulsive in the color--octet
channel and affects only the continuum states weakly.

\begin{figure}[t]
\begin{center}
\epsfig{file=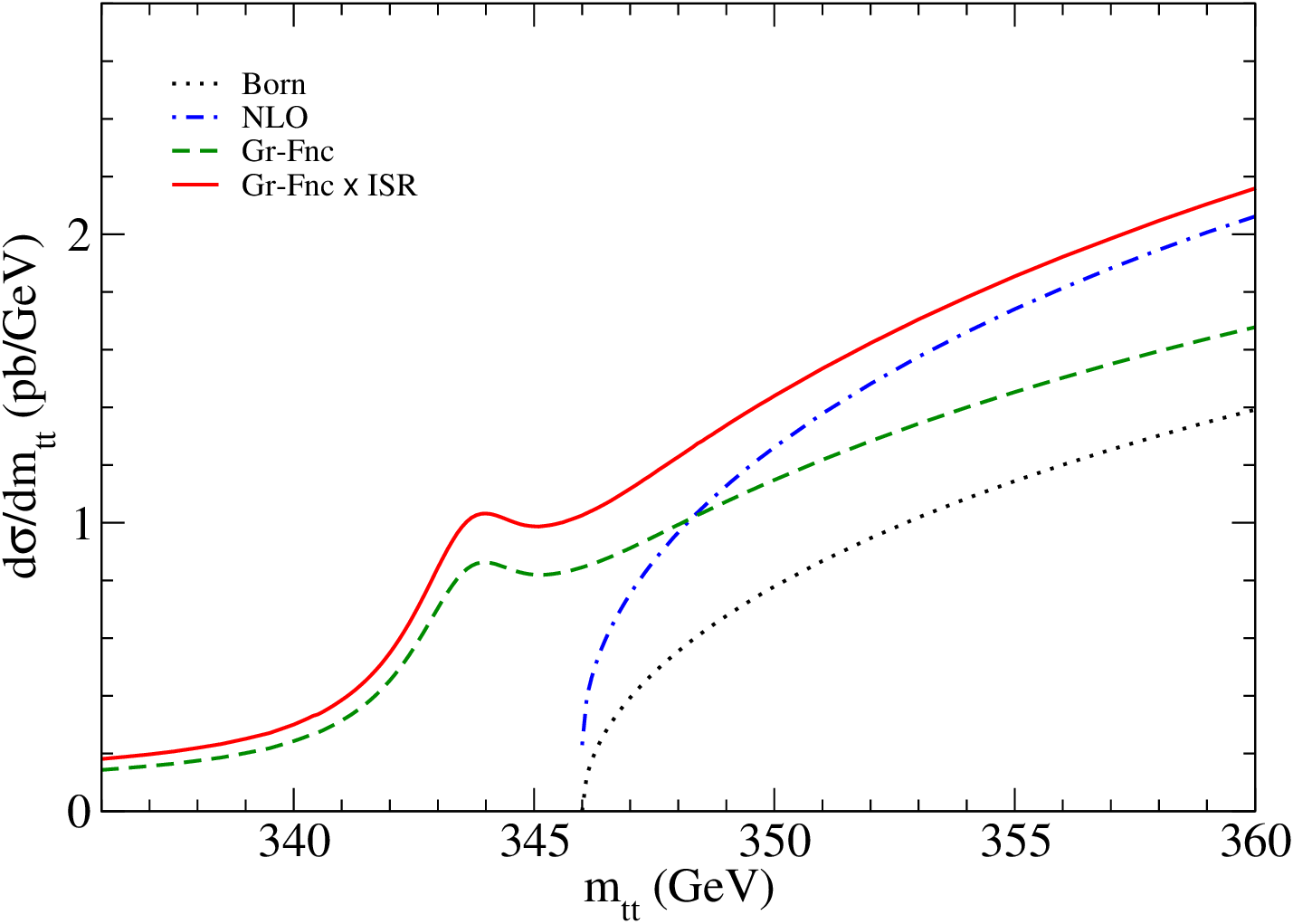,width=.6\textwidth}
\caption{
Same as Fig.~\ref{fig:sngj0} ($gg\to\ttb$), but the color and $J$ are
 summed.}\label{fig:gg}
\end{center}
\end{figure}
Fig.~\ref{fig:gg} shows a similar comparison of the $\ttb$
invariant-mass distributions
after the sum of the color--singlet and color--octet for all $J$
states are taken, for the $gg$ initial state.
They are obtained as the sum of the corresponding distributions in
Fig.~\ref{fig:sngj0} (color--singlet, $J=0$), those in Fig.~\ref{fig:oct}
(color--octet, all $J$), and the small contributions of $J\ge 2$ states
in the color--singlet channel, which are not shown separately.

\begin{figure}[t]
\begin{center}
\epsfig{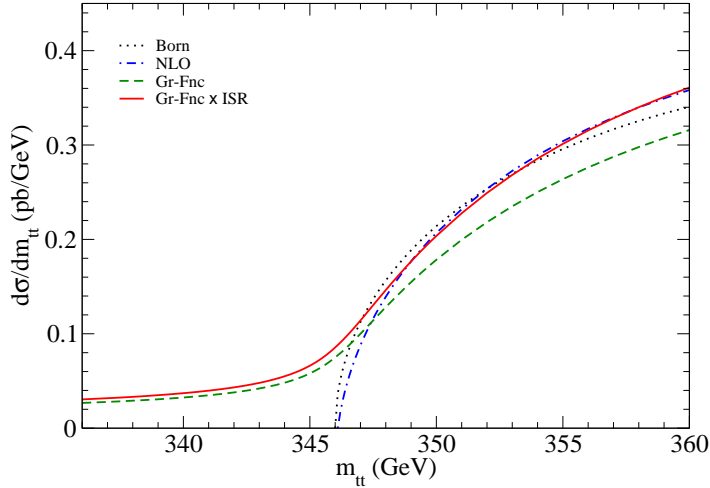}
\caption{
Same as Fig.~\ref{fig:gg}, but in $\qq$ channel (color--octet and
 $J=1$).}\label{fig:qq}
\end{center}
\end{figure}
Fig.~\ref{fig:qq} is for the $\qq$ initial state, where only the
color--octet $J=1$ channel exists.
Each distribution is similar to the corresponding one in the
color--octet $gg \to \ttb$ channel, see Fig.~\ref{fig:oct}. Note, however,
that the $\qq$ channel gets smaller enhancement by ISR than the $gg$
color--octet channel does
(the ratio of Gr--Fnc$\times$ISR and Gr--Fnc is about $1.1$ for the
$\qq$ channel, whereas the ratio is about $1.4$ for the color--octet $gg$
channel.).
The difference of the enhancement factors
originates from the different color factors in the ISR
functions, see Eqs.~(\ref{ISRfn}--\ref{subeq:k}).

\begin{figure}[t]
\begin{center}
\epsfig{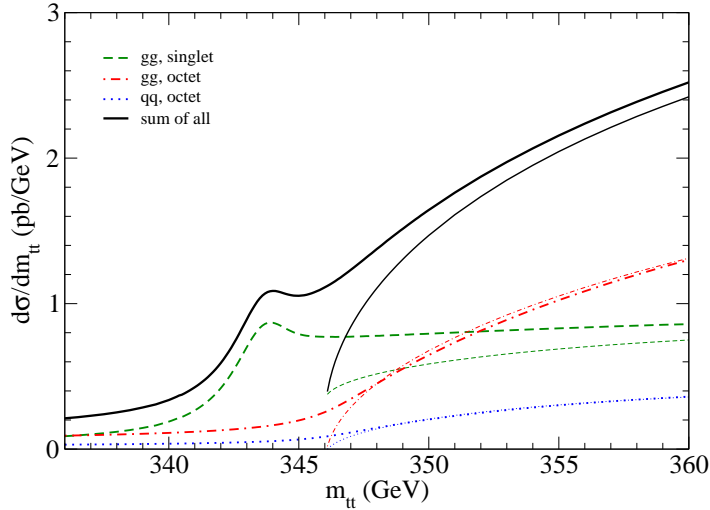}
\caption{
$\ttb$
invariant-mass distributions for the top-quark production 
 in the threshold region at LHC.
 The color--singlet (green dashed) and octet (red dot-dashed) in $gg$
 channel, color--octet in $\qq$ channel (blue dotted), 
and the sum of them (black solid) are plotted.
Thick lines include both bound-state and ISR effects, while thin lines
 represent the cross sections with only NLO effects.
}\label{fig:lhc}
\end{center}
\end{figure}
Displayed in Fig.~\ref{fig:lhc} are the invariant-mass 
distributions for $\ttb$ pair
 production at the LHC, where we show explicitly the individual
 contributions from the $gg$ color--singlet (dashed), $gg$ color--octet
 (dot-dashed), and $q\bar{q}$ (dotted) channels, as well as the sum of
 them (solid).
Thick lines include both bound-state and ISR effects
(Gr--Fnc$\times$ISR), while thin lines represent the NLO cross
sections. 
The invariant-mass distribution for the sum of all channels still
exhibits the $1S$ peak below the $\ttb$ threshold, while it gradually
approaches the NLO distribution above the threshold.
The color--singlet ($J$=0) $gg$ channel dominates the
cross section below and near the $\ttb$ threshold, while the
color--octet $gg$ channel is dominant above the threshold.
We expect that the cross section below the threshold\footnote{
The validity range of our formula is given by $|m_{\ttb}-2m_t|\simlt
\alpha_s^2 m_t$, which is of order 5--10~GeV.
}
($336~{\rm GeV} < m_{\ttb} < 2m_t = 346~{\rm GeV}$)
is about 6~pb, or $6\times 10^4$ events with 10~fb$^{-1}$.
The total enhancement due to the binding effects 
($-10~{\rm GeV} < m_{\ttb}-2m_t < 10~{\rm GeV}$) is estimated to be
about 8~pb, which corresponds to 1\% of the total NLO cross section.
It will be important to take them into account in detector-response
calibration and in the precise measurements of the top-quark mass. 

\begin{figure}[t]
\begin{center}
\epsfig{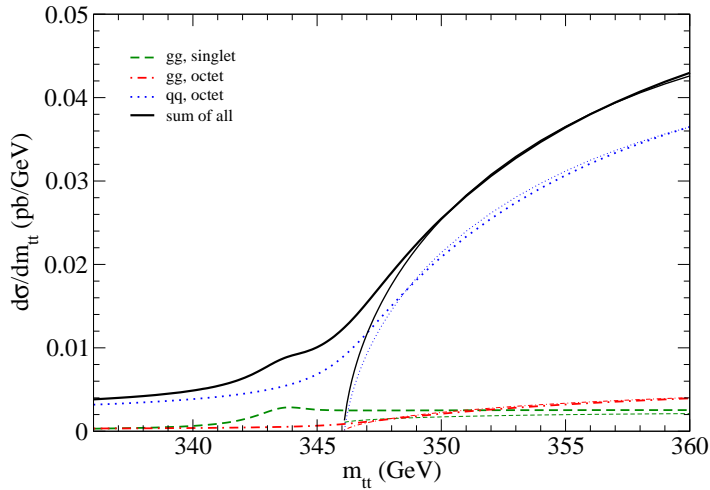}
 \caption{Same as Fig.~\ref{fig:lhc}, but for the cross section at
 Tevatron.}\label{fig:tev}
\end{center}
\end{figure}
Fig.~\ref{fig:tev} is the same as Fig.~\ref{fig:lhc} but for the
Tevatron, $p\bar{p}$ collision at $\sqrt{s}=1.96$ TeV.
Due to the dominance of $\qq$ channel which is color--octet in
the leading order, bound-state effects are less significant
in the threshold region.
The increase of the cross section {\it above} the threshold due to
bound-state effects,
as defined by the difference between Gr--Fnc$\times$ISR and NLO cross
sections at $0<m_{\ttb}-2m_t<10$~GeV,
is about 0.01 pb.
The ratio to the total cross section in NLO~\cite{Nason:1987xz} is
$2\times 10^{-3}$, which is consistent with the corresponding value
of Ref.~\cite{Catani:1996dj}.
However, the $\ttb$ pair-production cross section {\it below} the
threshold ($-10~{\rm GeV} < m_{\ttb}-2m_t<0$)
is estimated to be about 0.07~pb, or 70 events with 1~fb$^{-1}$.
Even though their contribution is still a small fraction, a few clean
events in this region could significantly affect the top-quark mass
determination at the Tevatron.

We may summarize the new aspects of our predictions
over the previous works as follows.
The most important difference of our predictions compared to
those of Ref.~\cite{Fadin:1990wx} is
the inclusion of ISR effects.
These effects increase the invariant-mass distributions
(Gr--Fnc)\
almost independently of $m_{\ttb}$ by about
20\%, 40\%, and 10\% for the
$gg\to\ttb$ color--singlet, color--octet, and $\qq\to\ttb$ color--octet
channels, respectively.
Furthermore, our predictions are more stable
against variations of $\mu_F$, $\mu_{\rm B}$ or $\mu_R$,
since we include the NLO corrections to bound-state effects
and ISR effects, while only LO bound-state effects are
incorporated in Ref.~\cite{Fadin:1990wx}.
Concerning the estimates of bound-state effects
on the $\ttb$ total cross sections in Ref.~\cite{Catani:1996dj},
we find that they are significant underestimates,
since only the effects above the threshold have been taken into
account.
\\

Summing up, we find that the bound-state effects significantly alter the
invariant-mass distributions of the $\ttb$ production close to
the threshold at the LHC.
The effects will be particularly important for the
determination of the top-quark mass \cite{Frederix:2007gi},
as well as when the top-quark sample is to be used for calibration of
jet energy scale, etc.\ at an early stage of the LHC operation.
In this regard, it should be noted
(known from the studies of $e^+e^-\to\ttb$ \cite{Sumino:1992ai})
that the bound-state effects distort the momentum distribution of
top quarks strongly below the threshold ($m_{\ttb}<2m_t$).
On the other hand, Tevatron experiments are almost free from the
bound-state effects, because of the dominance of the color--octet $\qq$
annihilation contribution, as can be seen from Fig.~\ref{fig:tev}.

We consider that our calculations incorporate the most important part of
the bound-state effects and ISR effects to the $\ttb$ threshold
cross sections at hadron colliders.
There are, however, many corrections that should be included to make
the predictions more precise.
Among them are resummation of the collinear and soft logarithms
(beyond NLO),
non-factorizable corrections, etc.
They are expected to modify our results at the level of 10--30\%.
Furthermore, as already stated, presently there remains a considerable
disagreement in a color-weighted sum
of the hard-vertex factors $h_{gg}^{(c)}$.
Hence, we examined the following case separately:
we use the value of $h_{gg}^{(1)}$ as in Eq.~(\ref{subeq:h}a),
which is determined from the two mutually consistent results
\cite{Kuhn:1992qw,Petrelli:1997ge};
on the other hand,
we determine the value of $h_{gg}^{(8)}$ such that the
$\alpha_s^3\beta$ term of the $gg\to\ttb$
cross section of Ref.~\cite{Nason:1987xz} is reproduced,
i.e.\ $h_{gg}^{(8)}(1)\approx 2.39$.
With this value of $h_{gg}^{(8)}$, the cross section in the
$gg$ color-octet channel is enhanced by about 10\%
as compared to the case with $h_{gg}^{(8)}(1)\approx -0.92$.
As a result, the cross section for
`sum of all' in Fig.~\ref{fig:lhc} becomes
more enhanced as $m_{\ttb}$ increases, where the enhancement
 is less than 2\% below the threshold, and is about 6\%
at $m_{\ttb}=360$~GeV.
By contrast, the cross section for
`sum of all' in Fig.~\ref{fig:tev} scarcely changes.
We also varied $\mu_R=\mu_F$ between $m_t/2$ and $2m_t$ and found that
the normalization of the cross sections changes by about 10\%
accordingly.
\\

%%%%%%%%%%%%%%%%%%%%%%%%%%%%%%%%%%%%%%%%%%%%%%%%%%%%%%%%%%%%%%%%%%%%%%%%%%%%%

% Acknowledgment
We are grateful to the referee for drawing our attention to
Ref.~\cite{Petrelli:1997ge}.
We are also grateful to M.~Czakon, A.~Mitov and 
the authors of Refs.~\cite{Nason:1987xz,Petrelli:1997ge}
(especially M.~Mangano, F.~Maltoni and P.~Nason)
for valuable communications.
The works of K.H.\ and Y.S.\ are supported in part by Grant-in-Aid for
scientific research (Nos.\ 17540281 and 17540228, respectively) from
MEXT, Japan.
We thank M.~Nojiri for organizing an LHC-focus week meeting at the IPMU
(Institute for the Physics and Mathematics of the Universe) in December
2007, where we enjoyed stimulating discussions.

% References

\end{document}